\documentclass[a4paper,11pt]{article}
\pdfoutput=1 
\usepackage{jcappub} 
\usepackage{amsfonts,graphics,color}
\usepackage{graphicx}
\usepackage{amsmath}
\usepackage[T1]{fontenc} 
\usepackage{float}
\usepackage{hyperref}
\usepackage{subcaption}
\usepackage{placeins}
\title{
Dynamical flattening of halo density cusps by Q-ball dark matter}

\author[a,1]{Alexander Libanov\note{Corresponding author.}}
\author[a,b]{and Sergey Troitsky}

\affiliation[a]{Institute for Nuclear Research of the Russian Academy of Sciences,\\ 60th October Anniversary Prospect 7a, Moscow 117312, Russia}
\affiliation[b]{Faculty of Physics, M.V. Lomonosov Moscow State University,\\ 1-2 Leninskie Gory,  Moscow 119991, Russia}

\emailAdd{amlibanov@inr.ac.ru}
\emailAdd{st@inr.ac.ru}

\abstract{
Cold, collisionless dark matter successfully explains a wide range of observations, including the formation of large-scale structure. Nevertheless, tensions remain on small, galactic scales, most notably the cusp–core, or inner-mass-deficit, problem and the diversity of inner rotation-curve shapes and central densities at fixed halo mass. These observations suggest that additional dark-sector physics may affect the inner structure of halos, although no generally accepted explanation has yet emerged.
Here, making use of a toy but representative model, we
investigate how interacting dark-matter Q-balls -- non-topological solitons stabilized by a conserved charge -- modify halo centers and generate a diversity of inner profiles. Our calculations include merger-induced mass and kinetic-energy losses, elastic heat transport, and the gravitational readjustment of the halo; flattening and later recontraction can both occur.
Produced in the early Universe in the dark sector, these Q-balls grow in the dense central regions of halos, while their interaction cross section decreases as the soliton mass increases. This process operates preferentially in halo centers, converting part of the rest-mass energy stored in massive Q-balls into relativistic dark-sector particles and thereby modifying the inner mass-density profile.
The resulting density-dependent energy loss competes with heat transport and halo response. An ensemble of initial assembly conditions yields very different final densities at fixed microphysics, although neither a universal constant-density core nor a statistical fit to the observed population is established.
}


\begin{document}
\maketitle

\flushbottom


\section{Introduction}
\label{sec:intro}
Small-scale structures in galactic halos remain challenging to reconcile with purely collisionless cold dark matter (CDM) models in the standard cosmological framework (for reviews see, e.g., \cite{cusp-core-rev1,cusp-core-rev2,cusp-core-rev3}). A notable example is the apparent absence of steep central density cusps in observed mass distributions \cite{cusp-core1,cusp-core2}, in tension with the cuspy inner profiles typically predicted in CDM-only models. Another tension arises from the diversity of galaxy halos, which show both cored and cusped profiles, see e.g. \cite{diversity1,diversity2,diversity3,diversity-redband}. Potential solutions include relaxing the assumption of collisionlessness: in particular, baryonic feedback and dark-matter self-interactions have been widely explored. However, each of these approaches faces its own challenges.

Baryonic feedback \cite{baryonic0,baryonic1,baryonic2} can smooth central dark-matter cusps under suitable conditions, for instance through repeated gas outflows that drive fluctuations of the gravitational potential. However, the inference of shallow inner density profiles in dark-matter-dominated low-surface-brightness galaxies and low-mass dwarfs \cite{few-baryons-LSB,little-THINGS} has raised concerns about whether feedback remains generically efficient in the most baryon-poor systems. Subsequent hydrodynamical simulations indicate that core formation depends sensitively on the timing and burstiness of star formation, often related to the stellar-to-halo mass ratio \cite{stellar-halo-ratio}, and may be further influenced by additional feedback channels, including black-hole activity in dwarfs \cite{not-full-solution-baryonic-diversity}. While these effects can contribute to halo-to-halo scatter, whether baryonic processes alone can account for the full diversity of inner halo profiles remains an unsettled question \cite{no-solution-diversity2019,not-full-solution-baryonic-diversity}.

On the other hand, self-interacting dark matter (SIDM) \cite{SpergelSteinhardt-SIDM} can naturally produce cored inner density profiles in dwarf galaxies (for reviews, see e.g.\ \cite{SIDM-review2017,SIDM-review2022}). The self-interaction cross section $\sigma$ per unit dark-matter particle mass, $m$, is tightly constrained on cluster scales \cite{Markevitch:2003at}. Since core formation in dwarfs typically requires a larger effective $\sigma/m$, SIDM models often invoke velocity-dependent scattering in the non-relativistic regime, with larger cross sections at dwarf-galaxy velocities than at cluster velocities. In addition, gravothermal evolution \cite{Gravothermal} can lead to distinct phases (core expansion and, for sufficiently strong interactions, core collapse), which may help account for the observed diversity of inner halo profiles \cite{diversity2019,gravothermal-diversity1,gravothermal-diversity2}. This explanation typically requires the effective cross section near the threshold for gravothermal collapse, so that only a subset of halos (e.g.\ the more concentrated ones) enter the collapse phase within a Hubble time, while others remain in the core-expansion regime \cite{diversity2019,collapse-in-MW-dwarfs}; it must also be consistent with constraints from black-hole growth and avoid producing overly massive central black holes through SIDM accretion \cite{SIDM-BH-growth}. Whether this mechanism can operate across the full population without tensions with other constraints remains an open question \cite{SIDM-review2022}.

The SIDM framework already points to environmentally dependent dark-matter phenomenology within halos, as is implied by the velocity dependence of the self-interaction cross section. This idea has been developed further in scenarios with segregated two-component dark matter \cite{segregation0}, where the heavier component preferentially sinks toward the halo center while the lighter component dominates the outskirts. Such constructions can help address several small-scale challenges of CDM \cite{segregation0,segregation1,segregation2}. In the present work, we pursue a different mechanism: the dark matter consists of macroscopic objects (Q-balls) whose effective properties evolve dynamically through mutual interactions. Because these interactions occur more frequently in the denser inner halo, the resulting evolution is inherently radius dependent. While a possible relation of Q-ball interactions to the halo structure was mentioned in Refs.~\cite{KusenkoSteinhardt,Sanchez-Salcedo:2005dru}, we are not aware of any detailed analysis of this relation and of its implications.

We outline our general framework in Sec.~\ref{sec:scenario:key} and introduce the toy model in Sec.~\ref{sec:scenario:model}. In Sec.~\ref{sec:calculation}, we formulate assumptions and derive equations for the evolving halo. Section~\ref{sec:constraints} discusses observational diagnostics and the limits of parameter constraints. In Sec.~\ref{sec:calculation-results}, we illustrate the model predictions with numerical examples. We summarize our conclusions in Sec.~\ref{sec:concl}.

\section{Scenario}
\label{sec:scenario}

\subsection[Overview]{Overview\protect\footnote{This subsection gives a sketch of the scenario we discuss here. See further sections for details and references.}}
\label{sec:scenario:key}
Scalar fields in a dark sector can form stable macroscopic nontopological solitons known as Q-balls. Their stability is ensured by conservation of a dark charge $Q$, and energetics typically favor a single large-$Q$ soliton over a collection of smaller-charge Q-balls or $Q$ free particles of unit charge. 
The mass relation makes coalescence energetically favorable, though by itself does not establish the fusion probability for moving Q-balls. 

If produced in the early Universe, Q-balls may constitute the dark matter; in this case, dark matter consists of many species with different charges, with masses and interaction cross sections that depend on $Q$. In the denser inner regions of halos, mergers are more frequent, driving the typical charge to larger values than in the outskirts. During these events, a fraction of the dark-matter rest-mass energy is released in the dark sector. 
We assume that the emitted quanta freely propagate away.

The resulting mass loss from the nonrelativistic component can generate an inner mass deficit. 
Whether this produces a core also depends on thermal transport and gravitational readjustment. The spread in final profiles arises from different assembly epochs and initial concentrations.

\subsection{The model}
\label{sec:scenario:model}
In this paper, we choose dark matter Q-balls as a model for self-interacting dark matter. A Q-ball is a non-topological soliton, the existence of which is allowed by many theories (for example \cite{FLS1, ColemanQB}). For calculations in this paper we use the Friedberg-Lee-Sirlin (FLS) theory of a real scalar field $\phi$ and a complex scalar field $\chi$ \cite{FLS1,FLS2}. The FLS Lagrangian, in notations of Ref. \cite{Troitsky:SMCO}, reads
\begin{equation}
    \label{FLSLagrangian}
       \mathcal{L} = \frac{1}{2}(\partial _{\mu} \varphi)^{2}-U(\varphi)+(\partial _{\mu}\chi)^{*}\partial_{\mu}\chi - k^{2}\varphi^{2}\chi^{*}\chi,
\end{equation}
\[
 U(\varphi)=(\varphi^{2}-v^{2})^{2},
\]
where $k$ and $v$ are some constants. 
The vacuum excitations have masses $m_\varphi=2\sqrt2\,v$ and $m_\chi=kv$ in these conventions. 
The model possesses a $U(1)$ symmetry rotating the field $\chi$, and hence the corresponding conserved Noether's charge $Q$. For a sufficiently large $Q$, the lowest-energy state is a spherical field configuration of radius $\sim R_Q$ with $\varphi \approx 0$ inside and $\varphi \approx v$ outside, called a Q-ball \cite{FLS1, FLS2}. 
For macroscopic Q-balls with a sufficiently large charge $Q$, the thin-wall approximation can be used. In this approximation, mass $M_{Q}$ and radius $R_{Q}$ of the solution with the charge $Q$ can be found by minimizing the classical energy derived from Eq.~(\ref{FLSLagrangian}), 
\begin{equation}
    \label{QballRadius}
    R_{Q}=\frac{1}{v}\left(\frac{Q}{4}\right)^{1/4},
\end{equation}
\begin{equation}
\label{QballMass}
    M_{Q} = \frac{4\sqrt{2}\pi}{3}vQ^{3/4}.
\end{equation}
Q-balls are stable with respect to decay into Q-charged particles provided
\begin{equation}
    \label{Qmin}
    Q\gtrsim Q_{\rm min}=\left(\frac{4\sqrt2\pi}{3k}\right)^4,
\end{equation}
which follows from  $M_Q/Q<m_\chi=kv$ and Eq.~(\ref{QballMass}). Throughout this work, we use the mass--radius relations of FLS solitons in the regime where the self-gravity of individual Q-balls can be neglected.

\section{Cosmological evolution of Q-balls}
\label{sec:calculation}

\subsection{Assumptions}
\label{sec:calculation:assumptions}
The toy model used in subsequent calculations is based on several assumptions.

We assume that Q-balls are born in the early Universe and comprise all dark matter. We stress the key difference with dark-matter supersymmetric Q-balls \cite{KusenkoShaposhnikov,KusenkoSteinhardt}: in our case, the solitons are formed in the dark sector (\ref{FLSLagrangian}) and do not interact with Standard-Model fields other than by gravity. We do not focus here on a particular mechanism of the Q-ball production, for which various options have been discussed in the literature, see e.g.\ Refs.~\cite{Krylov:QBbirth,McDonald:SuperheavyQballs} for dark-sector models. While Q-balls are produced with various charges, we assume that most of the solitons are born with charges close to a certain value $Q_{0}$, as suggested in a phase-transition production example \cite{Libanov:massgap}. 
Narrowness is an initial-condition approximation: mergers subsequently generate a charge distribution, whose effect we test for one representative halo.
We regard the value of $Q_0$ as a free starting parameter of the system, together with the parameters of the FLS Lagrangian (\ref{FLSLagrangian}).

After their birth, Q-balls form dark-matter halos by a standard mechanism. We assume that at sufficiently large redshifts, the number density of Q-balls is low enough, so we neglect their interactions and assume that the Navarro--Frenk--White (NFW) dark-matter density profile forms at some redshift $z_0$. In our toy model of the subsequent halo evolution, we consider it as an isolated system and do not study effects of subsequent halo mergers. 
The starting redshift $z_0$ is determined by mass-dependent effective initialization epochs from the assembly prescription of Refs.~\cite{Correa2015,Wechsler2002}, with a prescribed scatter in the epoch and concentration; see Eq.~(\ref{eq:ic-families}). Energy and mass can escape as dark radiation, so the bound halo is not a closed system.

After the halo is formed, we allow some Q-balls to interact and merge to produce populations of Q-balls with charges that depend on the distance $r$ from the halo center \cite{Libanov:massgap}. 
To calculate the time-dependent  radial density profile, we follow Lagrangian halo shells, including their mass, pressure and changing radii.

The rest energy of a Q-ball, Eq.~(\ref{QballMass}), implies that
$$
M_{Q_1+Q_2}<M_{Q_1}+M_{Q_2}.
$$
When two Q-balls merge, the expected energy is emitted in the form of relativistic particles and/or radiation \cite{Battye_2000, Heeck_2023, Ciurla_2024, Almumin_2022, Multamaki:50pcrosssection1, Multamaki:50pcrosssection2}. In our model, where fields of Eq.~(\ref{FLSLagrangian}) do not interact with massless vector fields, the emitted relativistic particles are associated with the real scalar field $\varphi$. Let us roughly estimate the fraction of energy carried away in $\varphi$, in thin-wall approximation \cite{FLS1}. The fraction of energy lost after merging of two Q-balls with identical charges $Q$ is
\begin{equation}
\label{phiEnergy2}
    \left( 2M(Q)-M(2Q) \right) /  \left( 2M(Q) \right)\approx 0.16.
\end{equation}
We assume that the energy released during relaxation is carried by escaping dark-sector scalar quanta. The total mass defect alone neither fixes the emitted spectrum nor guarantees escape. The recoil and kinetic-energy sink are accounted for in the evolution equations below.

Finally, we use a spherical, isotropic velocity distribution to characterize Q-ball encounters in the gravitational potential of the halo.
Our dynamical calculation evolves the one-dimensional velocity dispersion $s(r,t)$, distinct from the circular velocity $V_c(r)$, self-consistently with the changing mass and pressure profiles.

\subsection{Q-ball merging}
\label{sec:merging}
In this section, we discuss the effect of Q-ball merging on the dark-matter density in a particular halo. Within our assumptions, at the time of halo formation, Q-balls with identical charges $Q_{0}$ are the only component of dark matter, and each halo density profile is given by the NFW function,
\begin{equation}
    \label{NFWprofile}
        \rho_{\rm NFW}(r) = \frac{\rho_{N}}{\frac{r}{r_{N}}(1+\frac{r}{r_{N}})^{2}},
\end{equation}
where $r$ is the distance from the halo center, $r_{N}$ is the scale radius and $\rho_{N}$ is the density normalization \cite{NavarroProfile}.
We denote as $M_{200}(z)$ the mass enclosed within the radius at which the mean halo density is 200 times the critical density at redshift $z$. We relate $r_N$ and $\rho_N$ of the initial halo to its mass and concentration using {\sc COLOSSUS} \cite{COLOSSUS} and the {\sc Uchuu} simulations \cite{Uchuu}.
The concentration is independently varied around its reference value. We denote the nominal present-day halo mass used to specify the reference mass-accretion history as $M_{\rm label}$. The actual initial mass $M_{200}(z_0)$ is obtained from this history and generally differs from $M_{\rm label}$. Since we subsequently evolve the halo without cosmological accretion, $M_{\rm label}$ is not its calculated final mass either.

During subsequent halo evolution, Q-balls may merge or scatter elastically. Following the earlier geometric approximation, we distinguish
\begin{equation}
 \label{crosssection}
 \sigma_{\rm merge}=4\pi fR_Q^2,\qquad
 \sigma_{\rm el}=4\pi(1-f)R_Q^2,\qquad
 \sigma_{\rm tot}=4\pi R_Q^2.
\end{equation}
The illustrative constant fraction $f=0.5$ is motivated by phase-averaged simulations of other Q-ball models \cite{Multamaki:50pcrosssection1,Multamaki:50pcrosssection2}; it has not been calculated microscopically for FLS Q-balls. The velocity dependence of $f$ and possible partial charge transfer remain important uncertainties. If $f=f(g)$ depends on relative speed, the rate-weighted fraction is $f_{\rm eff}=\langle\sigma_{\rm tot}g f(g)\rangle/\langle\sigma_{\rm tot}g\rangle$; energy and heat-transfer moments need different weights. Our baseline keeps $f$ constant.

The Q-ball interaction strength is conveniently expressed through
\begin{equation}
 \label{eq:Xsigma}
 X\equiv vQ_0^{1/12},\qquad
 \left(\frac{\sigma_{\rm tot}}{m}\right)_0
 =\frac{3}{2\sqrt2}\,X^{-3}
 \simeq 0.1341\left(\frac{120\,{\rm MeV}}{X}\right)^3
 {\rm cm^2\,g^{-1}}.
\end{equation}
The first expression for the cross section uses natural units; $m=M_{Q_0}$. At fixed $X$, the absolute mass, radius, stability and microlensing properties still depend separately on $v$ and $Q_0$.

The initial halo density is given by Eq.~(\ref{NFWprofile}). Its initial potential determines the isotropic velocity dispersion; for a general spherical equilibrium profile,
\begin{equation}
 \label{dispersion}
 s^2(r)=\frac{1}{\rho(r)}\int_r^\infty \rho(r')
 \frac{GM(<r')}{r'^2}\,dr'.
\end{equation}
After an interaction step, the equilibrium profile must be recomputed. We model this response by hydrostatic equilibrium of Lagrangian shells,
\begin{equation}
 \label{eq:hydrostatic}
 \frac{dP}{dr}=-\frac{GM(<r)\rho}{r^2},\qquad P=\rho s^2,
 \qquad A\equiv\frac{s^2}{\rho^{2/3}},
\end{equation}
where $A$ is an entropy-like variable. Elastic encounters transport heat with luminosity
\begin{align}
 \label{eq:luminosity}
 L&=-4\pi r^2\kappa\frac{ds^2}{dr},\qquad
 \kappa^{-1}=\kappa_{\rm LMFP}^{-1}+\kappa_{\rm SMFP}^{-1},\\
 \kappa_{\rm LMFP}&=\frac{3aC}{8\pi G}\rho s^3\frac{\sigma_{\rm el}}m,
 \quad \kappa_{\rm SMFP}=\frac{3b}{2}\frac{s}{\sigma_{\rm el}/m},\nonumber
\end{align}
with $a=4/\sqrt\pi$, $b=25\sqrt\pi/32$ and baseline $C=0.84$. Here $C$ is a phenomenological conductivity coefficient, \emph{not} the NFW concentration or the constant of the former fixed-profile evolution equation. 
For the thermal conductivity $\kappa$, we adopt the interpolation between the long- and short-mean-free-path limits, $\kappa_{\rm LMFP, SMFP}$, commonly used in gravothermal models of elastic self-interacting dark matter (SIDM) \cite{Outmezguine:2022bhf,GadNasr2024,Palubski2024,Mace2026}. We apply this prescription to elastic Q-ball scattering as a phenomenological approximation.

For an initially narrow charge distribution, let $\mu=\Delta M_{\rm shell}(t)/\Delta M_{\rm shell}(0)$ denote the bound mass fraction of a Lagrangian shell. We approximate the evolving Q-ball population by a single effective charge $Q_{\rm eff}$. Assuming charge conservation within each shell and using $M(Q)\propto Q^{3/4}$, we obtain
\[
\frac{Q_{\rm eff}}{Q_0}=\mu^{-4}.
\]
Since $\sigma_{\rm merge}\propto Q^{1/2}$, this also gives
\[
\frac{\sigma_{\rm merge}}{m}
=\left(\frac{\sigma_{\rm merge}}{m}\right)_0\mu.
\]
Assuming independent local Maxwellian velocities with one-dimensional dispersion $s$, the mean relative speed is $\langle g\rangle=4s/\sqrt{\pi}$, and the merger rate per Q-ball is
\[
\Gamma_{\rm m}=\rho
\left(\frac{\sigma_{\rm merge}}{m}\right)_0
\mu\,\frac{4s}{\sqrt{\pi}}.
\]
Each binary merger is counted twice in the single-particle collision rate $\Gamma_{\rm m}$. For equal-charge Q-balls, $M(2Q)=2^{3/4}M(Q)$, which determines the loss of bound mass in mergers. Including the associated loss of random kinetic energy, we obtain the coupled gravothermal evolution equations
\begin{align}
\label{eq:gravothermal-sources}
\frac{d\ln\mu}{dt}
&=-\alpha\Gamma_{\rm m},
\qquad \alpha=1-2^{-1/4}\simeq0.15910,\\
\frac{d\ln A}{dt}
&=-\frac{2}{3s^2}\frac{\partial L}{\partial M}
-\beta\Gamma_{\rm m},
\qquad \beta=\frac76\,2^{-1/4}-\frac12\simeq0.48105.
\nonumber
\end{align}
Here $M=M(<r)$ denotes the enclosed bound mass, and $\beta$ accounts for the loss of random kinetic energy, including the effect of the decreasing shell mass. Its value is obtained assuming local Maxwellian velocities and isotropic scalar emission in the center-of-mass frame of each merging pair, with collisions weighted by their relative velocities. The emitted energy is treated as lost from the halo, as assumed above. The shell radii, masses and velocity dispersions evolve together with the gravitational potential. Our fluid description neglects charge-dependent velocity dispersions, baryons and cosmological accretion after initialization.

We convert effective initialization epochs to elapsed times using
a flat matter--$\Lambda$ cosmology,
\begin{equation}
 \label{z}
 t_{\rm age}(z)=
 \frac{2}{3H_0\sqrt{\Omega_\Lambda}}\,
 \operatorname{arsinh}
 \left[
 \sqrt{\frac{\Omega_\Lambda}{\Omega_m}}
 (1+z)^{-3/2}
 \right],
 \qquad
 t=t_{\rm age}(z)-t_{\rm age}(z_0),
\end{equation}
where $H_0=67.66\,{\rm km\,s^{-1}\,Mpc^{-1}}$ is the Hubble
constant, $\Omega_m=0.3111$, and $\Omega_\Lambda=1-\Omega_m$.

The initial assembly epoch is mass-dependent. We adopt a reference
$z_{\rm ref}$ from the Correa--Wechsler prescription
\cite{Correa2015,Wechsler2002}, with a reference Uchuu concentration,
and consider the following variations of the initial conditions,
\begin{equation}
 \label{eq:ic-families}
 \begin{aligned}
 1+z_0 &= a_z(1+z_{\rm ref}),
 &\qquad a_z &\in\{0.80,1,1.25\},\\
 c_0 &= c_{\rm ref}10^{\Delta\log_{10}c},
 &\qquad \Delta\log_{10}c &\in\{-0.12,0,+0.12\}.
 \end{aligned}
\end{equation}
These prescribed variations are not a fitted probability distribution. The reference concentration is retained upon shifting the epoch; the actual initialized mass $M_{200}(z_0)$ is supplied by the assembly prescription and generally differs from the mass label. No subsequent cosmological accretion or halo merger is included.

We solve the evolution equations numerically using 160 Lagrangian mass shells. The heat flux vanishes at both boundaries, while the outer pressure is kept at its initial value. The calculation describes the evolution of a halo after its formation, without subsequent cosmological accretion. Further details of the numerical implementation and accuracy tests are given in Appendix~\ref{app:numerics}.

\section{Parameters and solution}
\label{sec:constraints}
Before discussing representative results, we examine the constraints on our model and distinguish conditional diagnostics from observational parameter bounds.

\subsection{Q-balls as SIDM}
\label{sec:constraints:SIDM}
Self-interactions of dark matter are constrained by observations
of halos of different sizes, with the limits depending on the
characteristic particle velocities \cite{SIDM-review2022}.
For galactic and dwarf-galaxy halos, typical constraints are
relatively weak, $\bar\sigma_{\rm SIDM}\lesssim10$~cm$^2$/g
\cite{diversity2019,Galaxy_Crossection,Dwarf_Crossection}.
A more stringent constraint, conventionally expressed as
$\bar\sigma_{\rm SIDM}\lesssim1$~cm$^2$/g, comes from
observations of the Bullet Cluster at $z\simeq0.3$
\cite{Markevitch:2003at,BulletCluster}.

The Q-ball merger cross section per unit mass follows from
Eqs.~(\ref{QballRadius}), (\ref{QballMass}) and
(\ref{crosssection}),
\begin{equation}
 \label{meancrosssection}
 \bar\sigma_{Q,\rm merge}=\frac{\sigma_{\rm merge}}{M_Q}
 =\frac{3f}{2\sqrt2}v^{-3}Q^{-1/4}.
\end{equation}
Applying the conventional cluster-scale limit to the initial
total geometric cross section, $\bar\sigma_{Q, \rm tot}=\bar\sigma_{Q,\rm merge}/f$, gives $X\gtrsim60$~MeV. A detailed quantitative comparison with the Bullet Cluster would however require simulating the collision of two halos, including Q-ball mergers and the resulting mass loss, which is beyond the scope of the present paper. Here, we instead estimate the possible loss of dark matter from the smaller subcluster during a single passage through the main cluster.

Following earlier approximate analyses \cite{Feng2009}, we adopt a dark-matter column density $\mathcal S=0.30$~g\,cm$^{-2}$, a relative velocity $v_{\rm rel}=4800$~km\,s$^{-1}$, and an escape velocity $v_{\rm esc}=1200$~km\,s$^{-1}$ from the smaller subcluster. The latter determines whether a scattered Q-ball remains gravitationally bound. The adopted relative velocity is close to the observed shock velocity, which may exceed the actual relative velocity of the dark-matter halos \cite{SpringelFarrar}.

Assuming initially equal-charge Q-balls and isotropic elastic scattering, we estimate the fraction of mass lost from the smaller subcluster as
\begin{equation}
\label{eq}
f_{\rm loss}(X)=1-e^{-\tau_{\rm loss}(X)},\qquad
\tau_{\rm loss}(X)=\mathcal S\left[
\left(\frac{\sigma_{\rm merge}}m\right)_0+
0.875\left(\frac{\sigma_{\rm el}}m\right)_0\right].
\end{equation}
Here, the subscript 0 denotes cross sections evaluated at the initial Q-ball charge $Q_0$. The coefficient $0.875=1-2(v_{\rm esc}/v_{\rm rel})^2$ accounts for both ejection and capture in elastic collisions. For fixed merger and elastic scattering fractions, $\tau_{\rm loss}\propto X^{-3}$.

This estimate applies when interactions are sufficiently rare, which we require by imposing $\mathcal S(\sigma_{\rm tot}/m)_0\leq0.1$. For the adopted parameters, this corresponds to $X\gtrsim89$~MeV. Within this range, the estimated mass loss remains below approximately $10\%$; for example, $f_{\rm loss}\simeq3.7\%$ at $X=120$~MeV. These values are below the $\sim30\%$ mass loss conventionally considered in Bullet Cluster constraints \cite{Markevitch:2003at}. However, our estimate does not include the full dynamics of the cluster collision or the intracluster gas, and therefore cannot establish a firm observational bound on $X$. In particular, the condition $X\gtrsim89$~MeV limits the applicability of the approximation, not the allowed parameter space.

Merger-induced mass loss also acts during the secular evolution of a
cluster before any collision. We examine an isolated dark-matter halo
with initial $M_{200}=6.09\times10^{14}M_\odot$, concentration $c=1.68$,
and $(X,f,C)=(90~{\rm MeV},0.5,0.84)$, using 80 shells.
Over 6~Gyr, mergers remove $4.80\times10^{12}M_\odot$, or $0.135\%$
of the initial mass in the full computational domain extending to
approximately 18.8~Mpc. This fraction is not a mass loss within an
observational cluster aperture: in a fixed sphere of radius
$R\simeq950$~kpc, redistribution instead gives a net mass increase
of about $0.3\%$.
The sphere is defined by an initial mean dark-matter density of
$500\rho_{\rm crit,ref}$, where $\rho_{\rm crit,ref}$ follows from the
initial $M_{200}$ and $r_{200}$; its radius is a dark-matter-only proxy,
not the observational $r_{500}$ defined using the total mass.

To illustrate the possible effect on the gas fraction, we 
assign initial mass fractions of the gas,
$f_{\rm gas,0}=0.12$, and stars, $f_{\star,0}=0.02$, and  
hold the gas
and stellar masses inside this sphere fixed.
The calculated dark-matter redistribution then changes $f_{\rm gas}$
to approximately $0.1197$, a decrease of about $0.03\%$.
The sign of this small change depends on the aperture.
This conditional estimate illustrates the distinction between merger
mass loss and changes in enclosed mass; it does not model gas evolution
or establish agreement with observed cluster gas fractions.

\subsection{Q-balls as macroscopic dark matter}
\label{sec:constraints:macro}
Gravitational microlensing surveys, see e.g.\ Refs.~\cite{MACHO:1998, MACHO:2006, MACHO:2013, MACHO:2019}, constrain the abundance of compact dark objects in the Galactic halo. However, the usual constraints assume that the lenses can be approximated as point masses. We examine whether this assumption applies to the Q-balls considered here.

Microlensing probes objects distributed throughout the Galactic halo, where merger-induced charge growth is negligible in our scenario. We therefore consider Q-balls with their initial charge $Q_0$. Using $X=vQ_0^{1/12}$, Eqs.~(\ref{QballMass}) and (\ref{QballRadius}) give
\begin{align}
R_0 &= \frac{Q_0^{1/3}}{\sqrt{2}X}
\simeq 1.16\times10^6~{\rm km}
\left(\frac{120~{\rm MeV}}{X}\right)
\left(\frac{Q_0}{10^{72}}\right)^{1/3},
\\
M_0 &= \frac{4\sqrt{2}\pi}{3}XQ_0^{2/3}
\simeq 1.27\times10^{24}~{\rm g}
\left(\frac{X}{120~{\rm MeV}}\right)
\left(\frac{Q_0}{10^{72}}\right)^{2/3}.
\end{align}
The initial charge must satisfy the stability condition
$Q_0\gtrsim Q_{\min}$, Eq.~(\ref{Qmin}), but this condition does not
determine the absolute mass or radius of the Q-balls.

The numerical values illustrate how extended these objects
can be. For example, at $X=120$~MeV and $Q_0=10^{72}$,
a Q-ball has a mass of only $1.27\times10^{24}$~g but a
radius of about $1.7$ solar radii. At the same $X$, a
Q-ball with the mass of Jupiter would have a radius of
approximately $9.5$~AU, enormously larger than either
Jupiter itself or a black hole of the same mass.
These examples illustrate possible scales rather than
a predicted initial charge distribution.

For a lens of mass $M_0$, the Einstein radius is
\begin{equation}
R_E=\left(\frac{4GM_0D_{\rm eff}}{c^2}\right)^{1/2},
\qquad
D_{\rm eff}=\frac{D_l(D_s-D_l)}{D_s},
\end{equation}
where $D_l$ and $D_s$ are the distances to the lens and
source, respectively. Using Eq.~(3.4), we obtain
\begin{equation}
\frac{R_0}{R_E}\simeq 38
\left(\frac{120~{\rm MeV}}{X}\right)^{3/2}
\left(\frac{8~{\rm kpc}}{D_{\rm eff}}\right)^{1/2}.
\end{equation}
Remarkably, this ratio is independent of $Q_0$.
For a representative Galactic-halo lensing geometry toward
the Large Magellanic Cloud, $D_{\rm eff}=8$~kpc, the Q-ball
radius substantially exceeds the Einstein radius throughout
the range of $X$ considered in this work.

Consequently, conventional microlensing exclusions derived
for point-like compact halo objects cannot be directly
applied to these extended solitons. The magnification
produced by an extended Q-ball depends on its internal
mass-density profile, as discussed in Ref.~\cite{AnsariQBlensing}.
A quantitative constraint would require calculating the
lensing magnification for the FLS soliton profiles and the
corresponding survey detection efficiencies. We leave
such an analysis for future work and do not impose
additional constraints on $X$ or $Q_0$ from microlensing.

\subsection{Dwarf-galaxy checks}
\label{sec:constraints:summary}
\begin{figure}[tbp]
 \centering\includegraphics[width=0.85\linewidth]{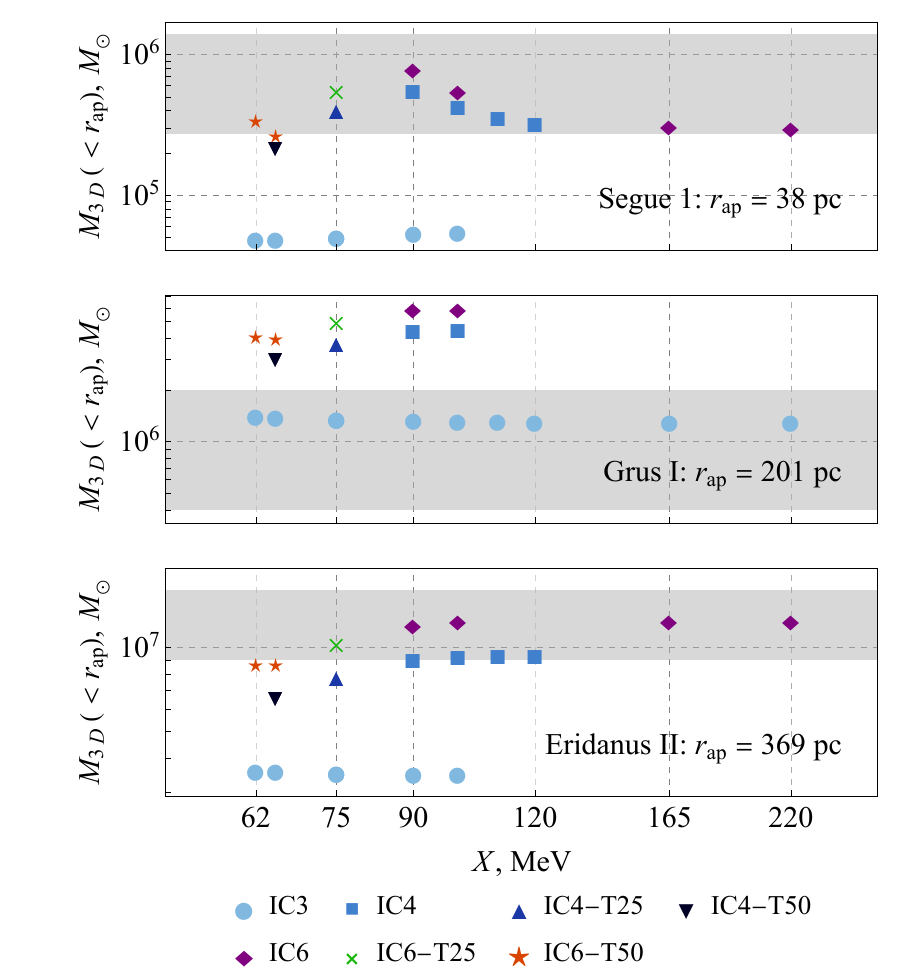}
 \caption{\label{fig:ufd-apertures}
 Final spherical dark-matter mass within the adopted three-dimensional
 half-light radius of Segue~1, Grus~I and Eridanus~II, evaluated
 for 54 aperture comparisons from 22 completed halo histories.
 The marker shapes identify the physical initial NFW halo models
 listed in Table~\ref{tab:ufd-initial-halos}; their symbols are
 the same in all three panels, but the matching halo model may
 differ between galaxies. The T25 and T50 suffixes denote
 intermediate assembly histories at fixed initial halo mass.
 Gray bands show the published asymmetric uncertainties on the
 enclosed dynamical mass
 \cite{Simon:2010seg,Chiti:2022grus,Li:2016eri}.
 For each of the seven sampled values $X=75$, 90, 100, 110,
 120, 165 and 220~MeV, at least one completed point lies within
 the gray band in \emph{every} panel, although the matching ICs
 need not coincide. This is an existence check for discrete
 parameter choices, not a joint confidence region or a
 continuous constraint on $X$.}
\end{figure}

We also compare the final spherical enclosed masses with
stellar-kinematic estimates for three ultra-faint dwarf galaxies.
The comparison is made at the three-dimensional half-light radius
$r_{1/2}$, rather than at the projected half-light radius $R_e$;
for dispersion-supported systems, the standard mass estimator gives
$M_{1/2}\simeq 3r_{1/2}\sigma_{\rm los}^2/G$ and
$r_{1/2}\simeq 4R_e/3$~\cite{Wolf:2010half}.
The corresponding
comparison bands are obtained directly from the published asymmetric
uncertainties on $M_{1/2}$:
$(2.7\text{--}14)\times10^5M_\odot$ for Segue~1
\cite{Simon:2010seg},
$(4\text{--}20)\times10^5M_\odot$ for Grus~I
\cite{Chiti:2022grus}, and
$(9\text{--}16)\times10^6M_\odot$ for Eridanus~II
\cite{Li:2016eri}. We adopt $r_{1/2}=38$, 201 and 369~pc for Segue~1, Grus~I and Eridanus~II, respectively. For Grus~I and Eridanus~II, the three-dimensional radii follow from the projected values $R_e=151$ and 277~pc used in the cited studies. The relevant consistency test is whether, at a fixed parameters of the model, $X$, $f$ and $C$, at least one completed halo model falls within the adopted mass interval \emph{for each} of the three galaxies. The halo models differ by initial conditions \eqref{eq:ic-families} and need not be identical for different galaxies. Figure~\ref{fig:ufd-apertures} demonstrates that this requirement is fulfilled at seven tested values $X=75$, 90, 100, 110, 120, 165 and 220~MeV in our finite sample of initial conditions, see Table~\ref{tab:ufd-initial-halos} in Appendix~\ref{app:numerics}. At $X=62$ and 65~MeV, no tested completed model reaches the nominal Eridanus~II interval, although the closest result at 65~MeV lies very close to it.

These discrete existence examples illustrate the freedom in
initial halo conditions at a common interaction strength;
they neither define an allowed interval in $X$
nor amount to a joint population fit. Matching one enclosed
mass does not establish a constant-density core or show that
Q-ball mergers are required. Stellar dynamics, tidal evolution
and the survival of the Eridanus~II star cluster are not modeled.

\begin{figure}[tbp]
\centering
\includegraphics[width=\linewidth]{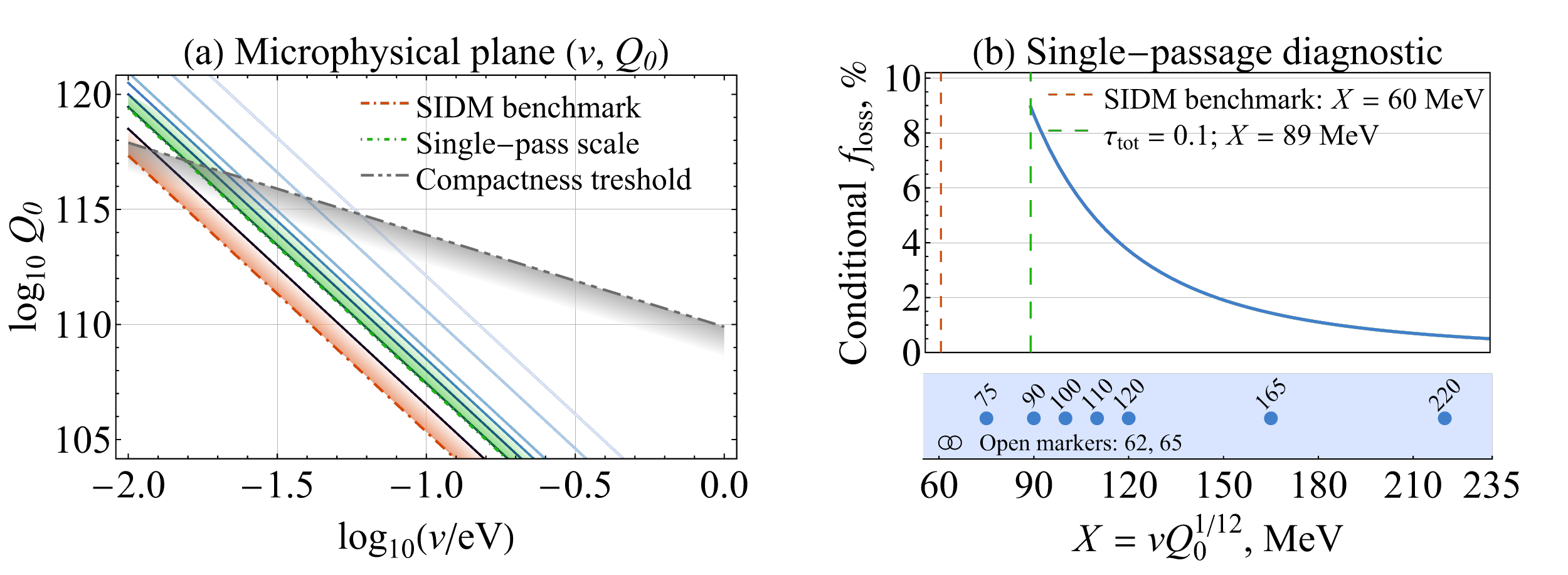}
\caption{\label{fig:parameter-overview}
The model parameter space and its constraints. 
(a) Solid contours show the seven tested values of $X=vQ_0^{1/12}$;
the other lines mark the conditional SIDM comparison, the
single-passage applicability scale and the compactness threshold
discussed in the text.
(b) Single-passage mass loss, Eq.~(\ref{eq}), for $f=0.5$ and the
collision parameters of Sec.~\ref{sec:constraints:SIDM}, restricted
to $\tau_{\rm tot}\leq0.1$.
Filled markers in the lower strip denote enclosed-mass matches for
all three dwarfs at $f=0.5$, $C=0.84$, allowing different initial
conditions for each galaxy. Open markers denote no Eridanus~II match
in the tested sample. }
\end{figure}

Figure~\ref{fig:parameter-overview} 
illustrates the range of discussed parameters in the $(v,Q_0)$ plane. It also presents a benchmark $R_0\gg2GM_0$, obtained from Eqs.~(\ref{QballRadius}) and
(\ref{QballMass}): to justify our calculations, soliton relations require negligible self-gravity. The stability threshold (\ref{Qmin}) lies
below the displayed charge range. The figure summarizes the distinct
roles of the conditional SIDM comparison, the single-passage estimate
and the discrete dwarf-galaxy checks.

\section{Halo evolution and diversity}
\label{sec:calculation-results}
To characterize the inner mass distribution of the evolving halos,
we use the ``cuspiness'' parameter introduced in
Ref.~\cite{diversity-redband},
\begin{align}
\Sigma &\equiv \frac{M_{\rm DM}(<0.01r_m)}{\pi(0.01r_m)^2},
\label{SigmaGeneral}\\
V_c^2(r)&=\frac{GM_{\rm DM}(<r)}{r},\qquad V_c(r_m)=V_{\max}.
\label{rotationcurve}
\end{align}
Here, $r_m$ is the radius at which the maximal circular velocity is reached. Both $r_m$ and $V_{\max}$ are determined from each final evolved halo profile, rather than assigned from the initial NFW halo. The quantity $\Sigma$ measures the enclosed central mass surface density but does not by itself distinguish a constant-density core from a shallower cusp.

\begin{figure}[tb]
 \centering\includegraphics[width=1.0\linewidth]{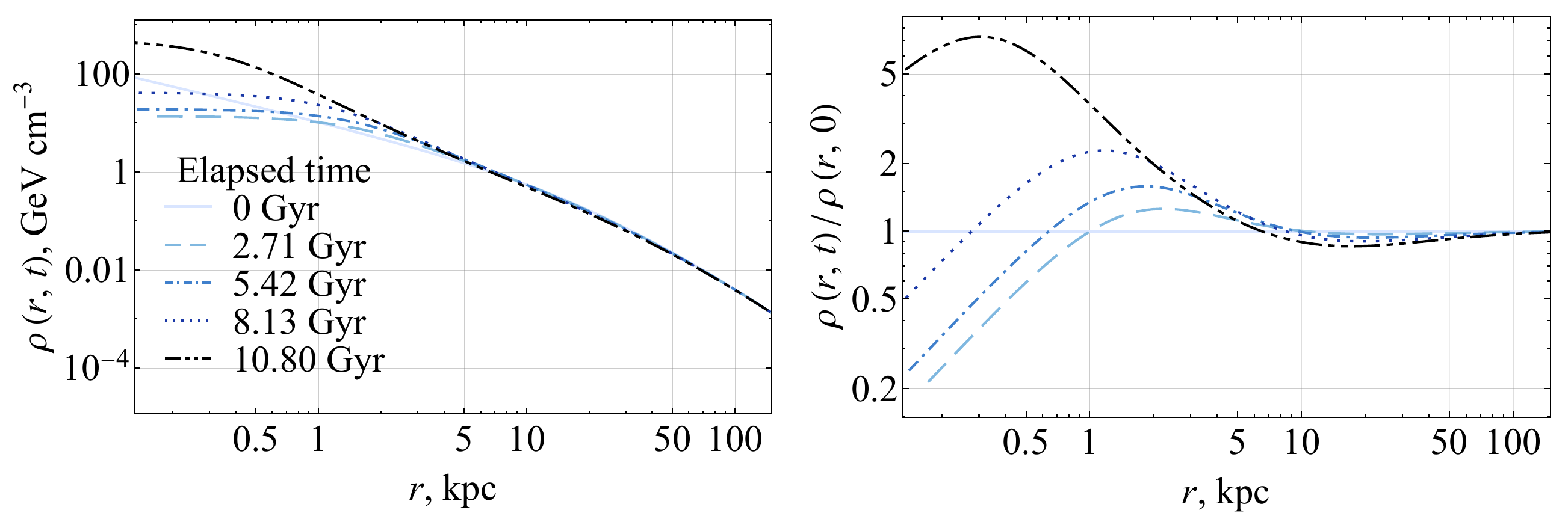}
 \caption{\label{fig:evolution} Density profiles at five elapsed times (left) and the ratio to the initial discrete density profile at the same physical radius (right) for one massive-halo history at $X=120$~MeV, $f=0.5$ and $C=0.84$. Each curve is restricted to the resolved radial domain.}
\end{figure}

As an illustration of the numerical evolution, we consider the reference set of physical parameters $(X,f,C)=(120~{\rm MeV},0.5,0.84)$. At the same microphysics, different initial assembly epochs and concentrations lead to different final profiles. Figure~\ref{fig:evolution} shows one example, with $M_{\rm label}=10^{13}M_\odot$, $a_z=1.25$ and $\Delta\log_{10}c=+0.12$. Its actual initialized mass is $M_{200}(z_0)=1.93\times10^{12}M_\odot$ at $z_0=2.219$. 
In this example, the inner density first decreases and
subsequently rises. This non-monotonic evolution reflects
the competition between merger-induced mass loss, which
weakens the central gravitational potential and favors
expansion, and the loss and redistribution of random
kinetic energy, which can reduce pressure support and
promote recontraction. The balance between these effects
changes as the merger rate, heat transport, and
gravitational structure of the halo evolve; see
Eqs.~(\ref{eq:hydrostatic})--(\ref{eq:gravothermal-sources}).
The late-time increase in density should not be
interpreted as a calculated terminal gravothermal
collapse, nor does an inner mass deficit necessarily
imply a constant-density core.

The main calculations employ the single-effective-charge
approximation described in Sec.~\ref{sec:merging}.
A comparison with an explicit charge-distribution calculation
for the halo of Fig.~\ref{fig:evolution} changes the final central
surface density by approximately 14\%.
We discuss this sensitivity test and its limitations in
Appendix~\ref{app:numerics}.

\begin{figure}[tb]
 \centering\includegraphics[width=1.0\linewidth]{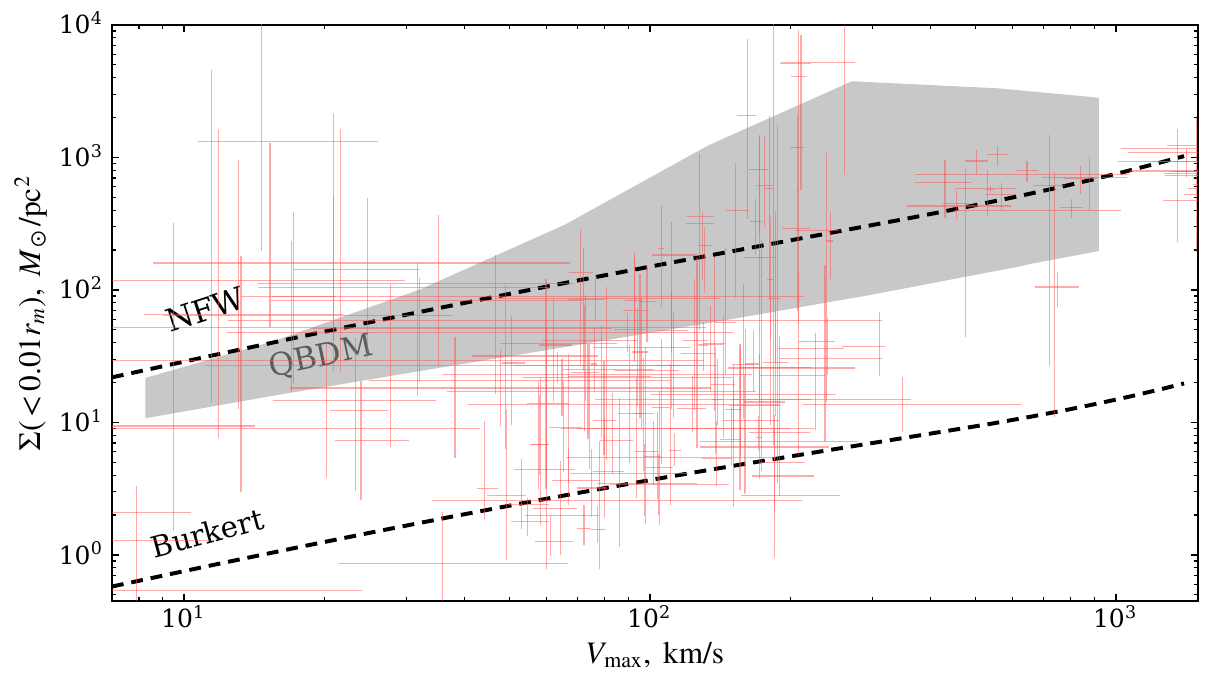}
 \caption{\label{fig:Cuspiness}The enclosed spherical dark-matter surface density $\Sigma$ versus final maximal circular velocity $V_{\max}$. Black reference curves show the NFW and Burkert relations. Red points show values \cite{diversity-redband} for observed halos and their uncertainties. The gray band covers nine prescribed initial-condition families \eqref{eq:ic-families}, computed from 72 completed halo histories at $X=120$~MeV, $f=0.5$, $C=0.84$.}
\end{figure}

To explore the diversity of halo profiles at fixed
microphysical parameters, we consider eight mass labels
between $10^8$ and $5\times10^{14}M_\odot$, each
combined with the nine assembly/concentration
configurations defined in Eq.~(\ref{eq:ic-families}).
All 72 resulting halo histories were evolved to
$z=0$ at fixed
$(X,f,C)=(120~{\rm MeV},0.5,0.84)$.
For each final mass distribution, we calculate
$V_{\max}$ using Eq.~(\ref{rotationcurve}),
evaluating the circular velocity at the resolved
shell-edge radii without extrapolation toward $r=0$.
Figure~\ref{fig:Cuspiness} shows the corresponding band over their common $V_{\max}$ interval (approximately 8.3--920~km\,s$^{-1}$) at fixed $X=120$~MeV. 
Observed values from Ref.~\cite{diversity-redband} are also shown. As phenomenological benchmarks, we use reference $\Sigma$--$V_{\max}$ relations for NFW and Burkert profiles, constructed from
the concentration--velocity parametrizations of
Refs.~\cite{SigmaVsVmax1,SigmaVsVmax2,diversity-redband}, see
Appendix~\ref{app:reference-relations}. This plot presents an example range over prescribed initial conditions and not an observational confidence region or a fitted distribution.

The NFW and Burkert reference relations help illustrate
the range of central densities, but do not delimit an
observationally allowed region. The comparison is not a
statistical fit to the observed population: the
initial-condition families are prescribed rather than sampled
from a calibrated distribution, and baryonic evolution is
neglected. 

Thus, even at fixed microphysical parameters, the final
central surface densities span a wide range as the initial
assembly epoch and concentration are varied. For the
illustrative set considered here, the spread in $\Sigma$
reaches factors of approximately 4, 15 and 40 at
$V_{\max}=30$, 100 and 300~km\,s$^{-1}$, respectively.
This demonstrates that Q-ball evolution can generate
substantial diversity in the inner mass distributions without
varying the microscopic interaction strength from halo to halo.

While the merger fraction $f=0.5$, motivated by numerical calculations in a different Q-ball model, is kept as a benchmark throughout the paper, we also test the sensitivity of the results to $f$ at $X=120$~MeV,
$C=0.84$ and $M_{\rm label}=10^{11}M_\odot$ for three histories:
late, median and early, with $(a_z,\Delta\log_{10}c)=(0.80,-0.12)$,
$(1,0)$ and $(1.25,+0.12)$, respectively. Using the same 160-shell
implementation, we obtain
$R_\Sigma\equiv\Sigma(z=0)/\Sigma(z_0)=(0.348,0.183,0.138)$
at $f=0.25$ and $(0.551,0.393,1.323)$ at $f=0.5$, respectively.
At $f=0.75$, the late and median histories
give $R_\Sigma=1.029$ and 1.932, while the early history reaches the
numerical density threshold specified in Appendix~\ref{app:numerics}
after 5.10~Gyr, indicating (but not establishing) the terminal physical collapse.

These examples show that varying $f$ can change both the magnitude and
the sign of the central surface-density response. At fixed $X$ and
charge, increasing $f$ enhances the merger cross section and reduces
the elastic cross section, but the resulting change in heat transport
depends on the local mean-free-path regime, Eq.~(\ref{eq:luminosity}).
This targeted comparison does not test the full diversity band or the
ultra-faint-dwarf matches at alternative $f$.
\section{Conclusions}
\label{sec:concl}
We have investigated the evolution of dark-matter halos
composed of interacting Q-balls, whose mergers convert
part of their rest mass into escaping dark-sector
radiation. The sublinear Q-ball mass--charge relation
leads to a loss of bound mass in mergers. Together with
the associated loss of random kinetic energy, elastic
heat conduction, and the gravitational readjustment of
the halo, this produces a nontrivial evolution of its
inner density profile. Depending on the initial halo
properties, the central density may decrease and
subsequently rise again. Thus, merger-induced mass loss
does not necessarily result in a monotonically expanding
core.

A central result is the diversity of halo structures
obtained at fixed microscopic parameters. For the
illustrative choice $X=vQ_0^{1/12}=120$~MeV, merger
fraction $f=0.5$, and conductivity coefficient $C=0.84$,
we evolved 72 halos spanning eight mass labels and nine
combinations of assembly epochs and initial
concentrations. The resulting spread in central surface
density $\Sigma$ at fixed $V_{\max}$ reaches factors of
approximately 4, 15, and 40 at
$V_{\max}=30$, 100, and 300~km\,s$^{-1}$,
respectively. This demonstrates how Q-ball interactions,
combined with differences in halo formation histories,
can produce a broad range of inner mass distributions
without varying the dark-matter microphysics between
halos.

We have also compared the calculated enclosed masses
with observational estimates for three ultra-faint
dwarf galaxies: Segue~1, Grus~I, and Eridanus~II.
At each of the seven tested values of $X$ between
75 and 220~MeV, we find initial halo configurations
that reproduce the adopted mass intervals for all
three galaxies, allowing their initial conditions
to differ while keeping the microscopic parameters
fixed. This provides an observational illustration
of the proposed diversity mechanism. The comparison
is an existence test over a discrete set of models,
rather than a statistical determination of the
allowed parameter range.

Our results indicate that Q-ball mergers, energy
transport, and halo assembly can jointly play an
important role in shaping the inner structure of
dark-matter halos. Quantitative population-level
predictions will require further development of the
interaction and gravothermal prescriptions, including
the evolution of charge distributions, as well as
more realistic halo assembly and baryonic effects.
Dedicated treatments of extended-object lensing and
cluster collisions will also be needed to establish
the corresponding constraints on the model.

\appendix
\section{Numerical implementation and supplementary tests}
\label{app:numerics}

\subsection{ Numerical implementation and accuracy checks}

The 72 halo histories used for the
$\Sigma$--$V_{\max}$ comparison in
Fig.~\ref{fig:Cuspiness} were calculated with
160 Lagrangian mass shells, initially extending
from $0.005r_N$ to $20r_N$.
The discretized NFW halo is first brought into
hydrostatic equilibrium. During its subsequent
evolution, the outer pressure is held fixed and
the heat flux vanishes at both boundaries.
The time step is limited to resolve mergers and
heat transport. Evolution is stopped if the
density of the innermost shell reaches 100
times its initial value; the 72 histories used
in the figure all reach the chosen final epoch.

Because mergers convert part of the Q-ball rest
mass into escaping radiation, and the outer
boundary can perform work, the bound mass and
thermal energy of the Q-ball component are not
separately conserved. A complete energy balance
must also include the emitted radiation and the
work done at the boundary. The kinetic-energy
term in Eq.~\eqref{eq:gravothermal-sources} uses
the collision-weighted moment
$\langle g^2\rangle_{\rm coll}=8s^2$ for
isotropic scalar emission in the pair
center-of-mass frame.

We tested the numerical stability of four
representative halos in seven additional
calculations. Four tests increased the resolution
from 160 to 240 shells, two halved the time
step, and one changed the innermost radial
boundary. The largest resulting differences
were about 0.013 dex in $\Sigma$ at fixed
aperture, $3.1\%$ in enclosed mass, and
$0.03\%$ in $V_{\max}$.
These checks support the numerical stability
of the selected results, but do not establish
convergence throughout the parameter space
or test the physical accuracy of the adopted
heat-conduction prescription. In particular,
their quantitative error estimates should
not be applied directly to the separate
ultra-faint-dwarf calculations, which use
80 Lagrangian shells.

\subsection{Ultra-faint-dwarf halo configurations}
The ultra-faint-dwarf comparison of
Sec.~\ref{sec:constraints:summary} uses seven
prescribed initial NFW halo configurations,
whose physical parameters are listed in
Table~\ref{tab:ufd-initial-halos}.
The three parent models are representative
early/high-density (IC4, IC6) and
late/low-density (IC3) initialization choices.
The additional T25 and T50 configurations
interpolate the assembly epoch and concentration
of IC4 and IC6 towards later, less concentrated
halos, while keeping their initial halo masses
fixed. Specifically, T25 and T50 stand for
interpolation fractions $t=0.25$ and $t=0.50$,
respectively, in the correlated prescription
\[
a_z(t)=1.25-0.45t,\qquad
\Delta\log_{10}c(t)=0.12-0.24t.
\]
The present-day assembly mass label is then
recalculated to preserve $M_{200}(z_0)$.
These are deliberately selected test cases,
not a probability sample of halo assembly
histories. The T25/T50 cases differ from the
nine discrete assembly/concentration variations
used for the $\Sigma$--$V_{\max}$ comparison.
\begin{table}[H]
\centering
\caption{
Physical initial conditions used in the
ultra-faint-dwarf comparison
(Fig.~\ref{fig:ufd-apertures}).
All rows describe the initial NFW profile
at $z_0$, before Q-ball evolution.
Here $c_0=r_{200}/r_s$, and $\rho_s$ is
the NFW density normalization.
T25 and T50 indicate intermediate assembly
histories, not different microphysics or
different evolution times.}
\label{tab:ufd-initial-halos}
\small
\begin{tabular}{lrrrrr}
\hline
Model & $M_{200}(z_0)$ & $z_0$ & $c_0$ & $r_s$ & $\rho_s$ \\
 & $(10^7M_\odot)$ & & & (kpc) &
 $(\mathrm{GeV\,cm}^{-3})$ \\
\hline
IC3       & 3.323 & 2.988 & 3.405 & 0.728 & 0.367 \\
IC4       & 4.123 & 4.973 & 5.819 & 0.308 & 3.999 \\
IC4-T25   & 4.123 & 4.484 & 5.085 & 0.383 & 2.276 \\
IC4-T50   & 4.123 & 3.984 & 4.444 & 0.482 & 1.266 \\
IC6       & 7.000 & 4.849 & 5.774 & 0.378 & 3.690 \\
IC6-T25   & 7.000 & 4.371 & 5.046 & 0.471 & 2.103 \\
IC6-T50   & 7.000 & 3.883 & 4.409 & 0.591 & 1.171 \\
\hline
\end{tabular}
\end{table}

For example, at $X=75$~MeV, the adopted mass intervals
for Segue~1, Grus~I, and Eridanus~II are reproduced
by IC4-T25, IC3, and IC6-T25, respectively.
At $X=120$~MeV, the corresponding configurations
are IC4, IC3, and IC4.
The same existence condition is satisfied at the
other five tested values of $X$ between 75 and
220~MeV. These discrete matches do not constitute
a population likelihood or demonstrate that
Q-ball mergers are required.

For the $X=120$~MeV examples, the inner
10-pc density ratios relative to each
run's own initial profile are 0.793 (IC3)
and 0.934 (IC4). Thus, agreement with
the observed enclosed masses does not
necessarily require strong central
density depletion. The comparison uses
only enclosed dark-matter masses;
stellar evolution, tidal effects, and
gas dynamics are not modeled.

\subsection{Sensitivity to the Q-ball charge distribution}

Successive mergers generally produce Q-balls with
different charges within the same Lagrangian shell.
To test the single-charge approximation, we therefore
follow the evolution of this charge distribution,
representing it numerically by a finite set of
charge groups. For each group $i$, we denote its
representative charge by $Q_i$ and the number of
Q-balls in the group by $N_i$. We define
$q_i=Q_i/Q_0$ and $c_i=N_i/N_0$, where $N_0$ is
the initial number of Q-balls in the shell, all
initially carrying charge $Q_0$.

Charge conservation and
$M(Q)\propto Q^{3/4}$ give
\begin{equation}
\label{eq:charge-moments}
\sum_i c_iq_i=1,\qquad
\mu=\sum_i c_i q_i^{3/4},\qquad
q_{\rm eff}=\mu^{-4}.
\end{equation}
The merger calculation includes interactions between all pairs of charge groups, allowing their geometric
cross sections to depend on both charges
and counting each pair only once.
Thus a single effective charge reproduces
the mass moment by construction, but need
not reproduce the merger and energy-loss
rates of a broad charge distribution.

\begin{figure}[tb]
\centering
\includegraphics[width=1.0\linewidth]
{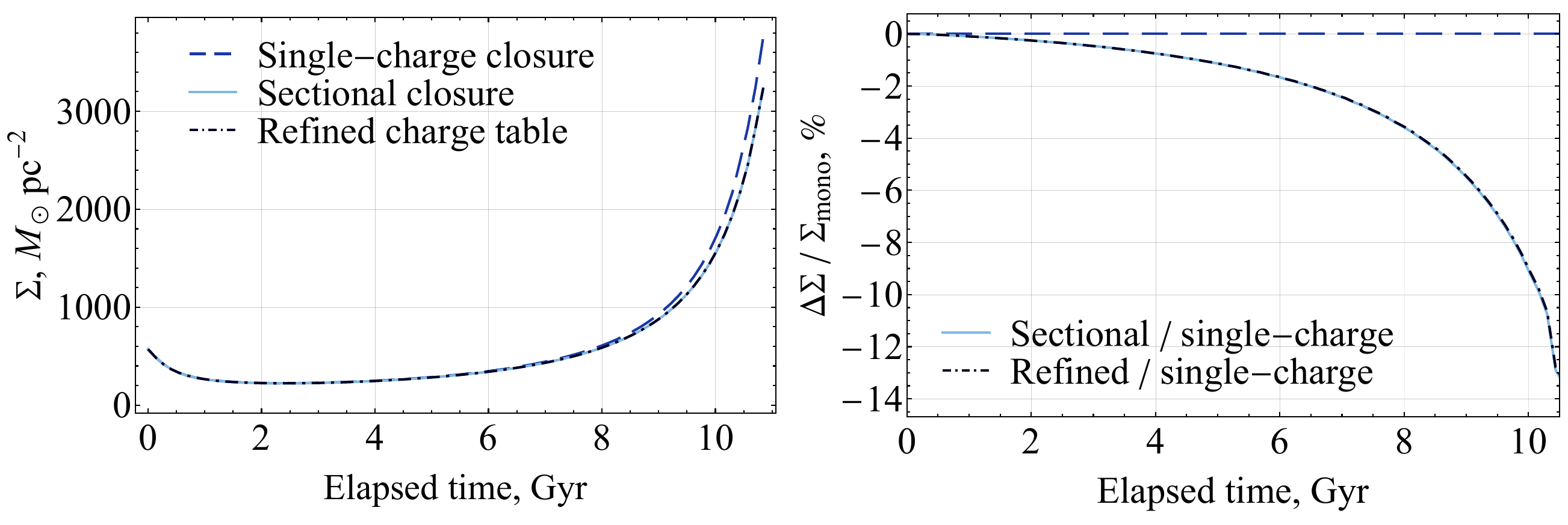}
\caption{\label{fig:charge-closure}
Effect of the Q-ball charge distribution
on the evolution of the same halo as in
Fig.~\ref{fig:evolution}, again for
$X=120$~MeV, $f=0.5$, and $C=0.84$.
Left: evolution of the central surface
density $\Sigma$ in the single-charge
approximation and with two discretizations
of the charge distribution.
Right: fractional differences relative
to the single-charge result.
This comparison tests the sensitivity
to the charge-distribution approximation
for one halo.}
\end{figure}

For the same initial halo as in
Fig.~\ref{fig:evolution}, we compared the
single-charge approximation with calculations
that explicitly evolve the charge distribution
in discrete bins.
Figure~\ref{fig:charge-closure} shows the
resulting evolution.
The final central surface density $\Sigma$
decreases from approximately $3756$ to
$3234\,M_\odot\,{\rm pc}^{-2}$, a difference
of $13.9\%$. Refining the charge grid gives
a similar difference of $13.8\%$.
Although non-negligible, this effect should
be considered in the context of other
approximations inherent in our
phenomenological treatment.
In particular, the charge-distribution
calculation still neglects charge-dependent
velocity distributions and spatial
segregation. We therefore leave a more
complete treatment of the evolution of
charge distributions to future work.

\section[Reference relations in the Sigma--Vmax plane]
{Reference relations in the $\Sigma$--$V_{\max}$ plane}
\label{app:reference-relations}

The NFW and Burkert curves in Fig.~\ref{fig:Cuspiness}
are reference relations at $z=0$, rather than
dynamical evolutions of individual halos.
Following
Refs.~\cite{SigmaVsVmax1,SigmaVsVmax2,diversity-redband},
we use the same functional form for both profiles,
\begin{align}
c_{V,P}(V_{\max})
&=c_{0,P}\left(1+\sum_{i=1}^{3}b_{i,P}y^i\right),
\label{eq:cv-reference}\\
\Sigma_P(V_{\max})
&=\frac{H_0V_{\max}\sqrt{c_{V,P}}}
{10^{-4}\sqrt{2}\pi G}
\frac{f_P(0.01x_P)}{f_P(x_P)},
\label{eq:sigma-reference}
\end{align}
where
\[
y=\log_{10}\frac{V_{\max}}{\mathrm{km\,s}^{-1}}.
\]
Here $P=N,B$ denotes the NFW and Burkert
profiles, respectively, and $c_{V,P}$ is the
mean density within the radius $r_{\max}$ of
maximum circular velocity, in units of the
critical density at $z=0$:
\[
c_{V,P}=\frac{\overline{\rho}(<r_{\max})}
{\rho_{\rm crit,0}}.
\]
The constants $x_P=r_{\max}/r_P$, where
$r_N$ and $r_B$ are the respective profile
scale radii, and the coefficients of the
concentration--velocity relations are given
in Table~\ref{tab:reference-profiles}.

\begin{table}[tb]
\centering
\caption{Parameters of the reference
$\Sigma$--$V_{\max}$ relations.
The NFW coefficients $b_{i,N}$ correspond
to $a_{i,N}^i$ in the original parametrization
of Ref.~\cite{SigmaVsVmax1}.}
\label{tab:reference-profiles}
\small
\begin{tabular}{lrrrrr}
\hline
Profile & $c_{0,P}$ & $b_{1,P}$ &
$b_{2,P}$ & $b_{3,P}$ & $x_P$ \\
\hline
NFW &
$7.21\times10^4$ & $-0.81$ &
$0.2209$ & $-0.019683$ & $2.162$ \\
Burkert &
$4.925\times10^4$ & $-0.8505$ &
$0.2434$ & $-0.02314$ & $3.24$ \\
\hline
\end{tabular}
\end{table}

The dimensionless enclosed-mass functions
are
\begin{align}
f_N(x)
&=\ln(1+x)-\frac{x}{1+x},
\label{eq:fN-reference}\\
f_B(x)
&=\frac14\ln(1+x^2)+\frac12\ln(1+x)
-\frac12\arctan(x).
\label{eq:fB-reference}
\end{align}
The NFW density profile is defined in
Eq.~\eqref{NFWprofile}, while the Burkert
profile \cite{Burkert:1995yz,SalucciBurkert}
is
\begin{equation}
\label{Burkert}
\rho_B(r)=\frac{\rho_{B,0}}
{(1+r/r_B)(1+r^2/r_B^2)},
\end{equation}
where $\rho_{B,0}$ is the central density
and $r_B$ is the core radius.

Both parametrizations are used over the
velocity range
$7\leq V_{\max}/(\mathrm{km\,s}^{-1})\leq1500$.
They provide reference curves for
Fig.~\ref{fig:Cuspiness} and do not
constitute observational bounds on the
Q-ball parameter $X$.

\acknowledgments
We are indebted to Dmitry Gorbunov, Maxim Libanov and Emin Nugaev for helpful discussions. This work was supported by the Russian Science Foundation grant 22-12-00215~(P).
\bibliography{qbdm}

\end{document}